\documentstyle[12pt,epsfig]{article}
\def\fe#1#2{
\begin{picture}(100,200)(0,0)
\put(34,0){\circle*{5}}
\put(0,35){\vector(1,-1){35}}
\put(-2,26){\makebox(0,0){#1}}
\put(-2,-26){\makebox(0,0){#2}}
\put(0,-35){\vector(1,1){35}}
\end{picture} }
\def\bp#1#2{
\begin{picture}(100,200)(0,0)
\multiput(0,0)(16,0){3}{\oval(8,8)[t]}
\multiput(8,0)(16,0){3}{\oval(8,8)[b]}
\put(#2,15){\makebox(0,0){#1}}
\end{picture}   }
\def\fpe#1{
\begin{picture}(100,200)(0,0)
\put(0,0){\line(1,0){35}}
\put(5,10){\makebox(0,0){#1}}
\end{picture}   }
\def\ft#1{
\begin{picture}(100,200)(0,0)
\put(0,0){\line(0,1){68}}
\put(10,24){\makebox(0,0){#1}}
\end{picture}   }
\def\hpe#1#2{
\begin{picture}(100,200)(0,0)
\multiput(0,0)(10.8,0){5}{\line(1,0){5.2}}
\put(#2,15){\makebox(0,0){#1}}
\end{picture}   }

\def\fspp#1{
\begin{picture}(100,200)(0,0)
\put(0,0){\circle*{5}}
\multiput(0,0)(14,-14){3}{\line(1,-1){10}}
\put(52,-36){\makebox(0,0){#1}}
\end{picture}   }

\def\bshaut#1{
\begin{picture}(100,200)(0,0)
\multiput(5,0)(8,8){5}{\oval(8,8)[tl]}
\multiput(5,8)(8,8){5}{\oval(8,8)[br]}
\put(34,30){\makebox(0,0){#1}}
\end{picture}   }

\begin{document}
\begin{flushright}
KEK Preprint TH-709 \\
August 2000 \\
\end{flushright}
\vspace{1cm}
\begin{center}
{\Large\sc {\bf Single Charged Higgs production as a probe of
CP violation at a Muon Collider}}
\vspace*{3mm}
{\large\sc {\bf }} 
\vspace{1cm}

{\large
{{ A.G. Akeroyd} and { S. Baek}} }
\vspace{1cm}
{\sl

KEK Theory Group, Tsukuba,\\
 Ibaraki 305-0801, Japan\\

\vspace{1cm}

}

\end{center}

\vspace{2cm}
\begin{abstract}
We consider single charged Higgs ($H^{\pm}$)
production in association with a $W^{\pm}$ boson
at $\mu^+\mu^-$ colliders, in the context of the general
CP violating Two Higgs Doublet Model (2HDM). We find that large 
cross-sections for the processes $\mu^+\mu^-\to H^+W^-,H^-W^+$ are possible,
and offer an attractive way of producing $H^{\pm}$ at $\mu^+\mu^-$ colliders.
The difference in the cross-sections for $H^+W^-$ and $H^-W^+$ may 
exceed 1000 fb, and this represents a novel way of probing CP violation
in the Higgs sector.

\end{abstract}

\newpage
\section{Introduction}
Charged Higgs bosons ($H^{\pm}$) are predicted in many 
extensions of the Standard Model (SM), in particular the Minimal Supersymmetric
Standard Model (MSSM). Their phenomenology \cite{Gun} has received 
much attention both at $e^+e^-$ colliders \cite{e+e-}
and at hadron colliders \cite{Hadron}, \cite{recent},\cite{progress}. 
At $e^+e^-$ colliders production proceeds via the
mechanism $e^+e^-\to \gamma^*,Z^*\to H^+H^-$,
with higher order corrections
evaluated in \cite{eeHH}, and detection is possible for
$M_{H^{\pm}}$ up to approximately $\sqrt s/2$. The combined
null--searches from all four LEP collaborations derive the lower
limit $M_{H^{\pm}}\ge 77.3$ GeV $(95\%\, c.l)$ \cite{LEP}. 

In recent years an increasing amount of work has been dedicated to 
the physics possibilities of $\mu^+\mu^-$ colliders 
\cite{Muon}, \cite{talks}. Such colliders offer novel ways of 
producing Higgs bosons, and much attention has been given to the
study of neutral Higgs bosons produced as resonances in the s-channel
\cite{res},\cite{muon}.

The phenomenology of $H^{\pm}$ at $\mu^+\mu^-$ colliders
has previously been considered to be more or less identical to that at 
$e^+e^-$ colliders. This is because the pair production processes
of $e^+e^-,\mu^+\mu^-\to H^+H^-$ have been assumed to have the same rate
at both colliders. This is the case in the MSSM, where
the Higgs mediated s-channel diagrams have been shown to be negligible
at a $\mu^+\mu^-$ collider
\cite{AAD}. The single production of $H^{\pm}$ via the process 
$e^+e^- \to H^{\pm}W^{\mp}$ \cite{eehw}, which proceeds dominantly via loops,
has relatively small rates. An analysis in the context of the LHC
has been covered in \cite{pphw}. At a muon collider
this process can have a much larger cross-section because the 
tree-level diagrams, 
which are suppressed by $m^2_e$ in the $e^+e^-$ case, are 
proportional to $m^2_{\mu}$, and become by far the dominant contribution.

The mechanism $\mu^+\mu^-\to H^{\pm}W^{\mp}$
was first considered in \cite{Alan} and subsequently developed in \cite{AAD}. 
It possesses several advantages over
the conventional pair production process, $\mu^+\mu^-\to H^+H^-$.
In particular, $H^\pm$ may be produced on-shell for 
$M_{H^\pm}\le \sqrt s-M_W$, which compares favourably 
with the kinematic reach for pair production ($M_{H^\pm}\le \sqrt s/2$). 
In addition, backgrounds are expected to be relatively small, since
for $H^{\pm}\to tb$ decays the main background would be from 
$\mu^+\mu^-\to t\overline t$ production which has a cross-section
of 700 fb at $\sqrt s=500$ GeV.
In \cite{AAD} an analysis in the context of the MSSM showed that 
sizeable cross-sections ($\ge 20$ fb) can be attained for 
$\tan\beta\ge 40$. In this paper we consider
the general (non-SUSY) Two Higgs Doublet Model (2HDM), 
which has the added advantage of allowing CP-violation in the
tree-level Higgs potential. 
In contrast to the MSSM, all the Higgs masses may be taken 
free parameters and so one would expect
larger cross-sections, as well as CP asymmetries in the rates for 
$\mu^+\mu^-\to H^+W^-,H^-W^+$. We will show that such a production 
mechanism may provide a copious source of $H^{\pm}$ as well as offering
a novel way of probing CP violation in the Higgs sector, the latter not being
possible in the standard mechanism $\mu^+\mu^-\to H^+H^-$.

Our work is organized 
as follows. In Section 2 we introduce the 2HDM potential, and section 3
derives explicit formulae for the cross-sections. 
In Section 4 we present our numerical analysis while section 5 
contains our conclusions.

\section{2HDM Potential}
The most general 2HDM potential which violates CP and
only softly breaks (by dimension 2 terms) 
the discrete symmetry $\Phi_{i}\to -\Phi_{i}$
contains 8 free parameters at tree-level \cite{Lee}. 
We will follow the notation of \cite{GGK}. 
The potential is given as follows.
\begin{equation}
V(\Phi_{1}, \Phi_{2})=V_{symm}+V_{soft}
\end{equation}
where
\begin{eqnarray}
 V_{symm} =  -\mu^2_{1}\Phi^{\dagger}_{1}\Phi_1
-\mu^2_{2}\Phi^{\dagger}_{2}\Phi_2+
\lambda_1(\Phi^{\dagger}_{1}\Phi_1)^2+ 
\lambda_2(\Phi^{\dagger}_{2}\Phi_2)^2+ \nonumber \\
\lambda_3(\Phi^{\dagger}_{1}\Phi_1)(\Phi^{\dagger}_{2}\Phi_2)
+\lambda_4|\Phi^{\dagger}_1\Phi_2|^2+
{1\over 2}[\lambda_5(\Phi^{\dagger}_1\Phi_2)^2+h.c]
\end{eqnarray}
and
\begin{equation}
V_{soft}=-\mu^2_{12}\Phi_1^{\dagger}\Phi_2+h.c
\end{equation}
CP violation, either spontaneous or explicit, requires the
presence of $V_{soft}$ which breaks the discrete symmetry softly. 
If all the parameters are real, spontaneous CP violation can still
occur provided that:
\begin{equation}
\left|{\mu_{12}^2\over \lambda_5v_1v_2}\right|\le 1.
\end{equation}
CP violation will be explicit if Im($\mu^{*4}_{12}\lambda_5\ne 0$). 
In the CP conserving case one finds two CP even neutral scalar eigenstates,
$h^0,H^0$, and a CP odd eigenstate $A^0$. 
In the CP violating case, mixing is induced between the CP even
and CP odd neutral scalar fields, resulting in three mass eigenstates
$H_1,H_2,H_3$ with no definite CP quantum numbers. In the MSSM, such mixing
may be induced when one considers the 1-loop effective scalar potential.
This would also lead to a rate asymmetry in the processes
$\mu^+\mu^-\to H^+W^-,H^-W^+$ and this will be addressed in \cite{akebaek}.
The neutral scalar mass squared matrix ${\cal M}^2_S$ is diagonalized by the 
matrix $O_{ij}$:
\begin{equation}
O^T{\cal M}^2_SO=diag( M^2_{H_1},M^2_{H_2},M^2_{H_3})
\end{equation}
We will parametrize the matrix $O_{ij}$ by using three Euler angles
as follows.

\begin{equation}
O_{ij}=\left[\matrix{
c_{12}c_{13} & s_{12}s_{13} & s_{13} \cr
-s_{12}c_{23}-c_{12}s_{23}s_{13} & c_{12}c_{23}-s_{12}s_{23}
s_{13} & s_{23}c_{13} \cr
s_{12}s_{23}-c_{12}c_{23}s_{13} & -c_{12}s_{23}-s_{12}c_{23}
s_{13} & c_{23}c_{13}} \right]
\end{equation}

The CP conserving limit is obtained by taking
two of the Euler angles equal to zero, and so the 
eigenstates of ${\cal M}^2_S$ become pure CP eigenstates,
$h^0,H^0$ and $A^0$. This results in a potential with 6 free parameters,
$V_{symm}$. The condition for maximum CP violation was considered in \cite{Pom}.

\newpage
\section{$\mu^+\mu^-\to H^{\pm}W^{\mp}$}
\noindent
%
\par
\begin{center}
\begin{picture}(100,100)(0,0)
\put(-40,69){\vector(1,-1){1}}
\put(-40,31){\vector(-1,-1){1}}
\put(-60,50) {\fe{$\mu^-$}{$\mu^+$}}
\put(-24,50) {\hpe{$H_1,H_2,H_3$}{24}}
\put(25,50) {\bshaut{$\ \ \qquad W^-$} }
\put(23,50) {\fspp{$H^+ $}}
\put(129,90){\vector(1,0){1}}
\put(129,20){\vector(-1,0){1}}
\put(150,55){\vector(0,-1){1}}
\put(110,90) {\fpe{$\mu^-$}}
\put(150,90){\circle*{5}}
\put(146,21) {\ft{$\ $}}
\put(135,49) {$\nu_\mu$}
\put(150,21){\circle*{5}}
\put(110,20) {\fpe{$\mu^+$}}
\put(150,90) {\bp{$\ $}{20}}
\put(152,21) {\hpe{$\ $}{30}}
\put(185,93) {$W^-$}
\put(185,29) {$H^+$}
\put(42,-10) {{\bf Figure.1}}
\end{picture}
\end{center}
\vspace{0.6cm}
Single $H^{\pm}$ production may proceed via an s--channel resonance 
mediated by
$H_i$, and by $t$-channel exchange of $\nu_{\mu}$ 
(see Fig.~1). We will present explicit formulae for 
the processes $\mu^+\mu^-\to H^+W^-$ and $\mu^+\mu^-\to H^-W^+$ by
adapting the formulae presented in \cite{AAD}, to which we refer the reader
for a detailed explanation of our notation. As explained in \cite{AAD},
model II type couplings are required for this production mechanism
to have an observable rate. 

The CP violation originates
from the $s$-channel diagrams and the $st$ interference, and 
is caused by the elements
of $O_{ij}$ which mix the pure CP even and CP odd scalar fields. 
In the $s$-channel diagrams 
the couplings at the vertices ($g_{H_iH^+W^-},g_{H_i \bar{\mu} \mu}$), 
which are either purely real or purely
imaginary in the CP conserving case, possess both a real and imaginary part.
We will show that this induces a difference in the rates for 
$\mu^+\mu^-\to H^+W^-$ and $\mu^+\mu^-\to H^-W^+$. 
The CP violating couplings are as follows:
\begin{eqnarray}
g_{H_i H^\pm W^\mp}&:&  (O_{2i}\cos\beta- O_{1i}\sin\beta,\,\, O_{3i}) 
\nonumber \\
g_{H_i \bar{\mu} \mu}&:&  (O_{1i},\,\, O_{3i} \sin\beta),
\end{eqnarray}
where $i=1,2,3$.  

We now present the formulae for the matrix elements for 
for $H^+W^-$ and $H^-W^+$ production.
The matrix element squared for $\mu^+\mu^-\to H^{+}W^{-}$ is as follows:
\begin{eqnarray}
&&|{\cal M}|^2(\mu^+\mu^-\to H^{+}W^{-}) 
=\frac{s g^4 m_{\mu}^2 }{32M_W^4} \Bigg[\nonumber \\
&&{\lambda( s,M_{H^\pm}^2,M_W^2)\over\cos^2\beta}
\sum_{i,j}
g_{H_i H^\pm W^\mp} g_{H_j H^\pm W^\mp}^* S_{H_i} S^*_{H_j}
{\rm Re} \Big\{g_{H_i \bar{\mu} \mu} g_{H_j \bar{\mu} \mu}^* \Big\}
 \nonumber \\
&+& 2\tan^2\beta S_F^2(t) (2 M_W^2 p_T^2 + t^2 ) \nonumber \\
&+& {\tan\beta\over \cos\beta} 
S_F(t)(M_{H^\pm}^2 M_W^2-s p_T^2-t^2)\sum_{i} \Big\{g_{h_i H^\pm W^\mp}
g_{H_j \bar{\mu} \mu}
S_{H_i}+ c.c \Big\}
\Bigg]
\label{HpWm}
\end{eqnarray}
Where $p^2_T=\lambda(s,M_{H^{\pm}}^2,M_W^2) \sin^2\theta/4s$,
$S_F(t)=1/t$, and the propagators $S_{H_i}$ are given by:
\begin{equation}
S_{H_i}= {1\over s-M_{H_i}^2+
iM_{H_i}\Gamma_{H_i}}
\end{equation}
The matrix element squared for $\mu^+\mu^-\to H^-W^+$:
\begin{eqnarray}
&&|{\cal M}|^2(\mu^+\mu^-\to H^{+}W^{-}) 
=\frac{s g^4 m_{\mu}^2 }{32M_W^4} \Bigg[\nonumber \\
&&{\lambda( s,M_{H^\pm}^2,M_W^2)\over\cos^2\beta}
\sum_{i,j}
g_{H_i H^\pm W^\mp}^* g_{H_j H^\pm W^\mp} S_{H_i} S^*_{H_j}
{\rm Re} \Big\{g_{H_i \bar{\mu} \mu}^* g_{H_j \bar{\mu} \mu} \Big\}
 \nonumber \\
&+& 2\tan^2\beta S_F^2(t)(2 M_W^2 p_T^2 + t^2 ) \nonumber \\
&+& {\tan\beta\over \cos\beta} 
S_F(t)(M_{H^\pm}^2 M_W^2-s p_T^2-t^2)\sum_{i} \Big\{g_{h_i H^\pm W^\mp}^*
g_{H_i \bar{\mu} \mu}^*
S_{H_i}+ c.c \Big\}
\Bigg]
\label{HmWp}
\end{eqnarray}
The origin of the CP violation 
is the interference between the weak phases 
(phases in the $g_{H_i H^\pm W^\mp}$ and $g_{H_i \bar{\mu} \mu}$)
and absorptive phases (phases in the $S_{H_i}$),
as can be seen in (\ref{HpWm}) and (\ref{HmWp}).

The differential cross--section for 
$\sigma(\mu^+\mu^-\to H^{\pm}W^{\mp})$ may be written as follows:
\begin{equation}
{d\sigma\over d\Omega} = {\lambda^{1\over 2}(s,M_{H^{\pm}}^2,M_W^2)\over
64\pi^2s^2} |{\cal M}|^2 \label{ref1}
\end{equation}
The total cross-section, $\sigma_{tot}$, is defined by:
\begin{equation}
\sigma_{tot}=\sigma(\mu^+\mu^-\to H^+W^-)+\sigma(\mu^+\mu^-\to H^-W^+)
\end{equation}
In the CP conserving case the $g_{H_i H^\pm W^\mp}$ are $g_{H_i \bar{\mu} \mu}$ 
are either purely real or purely imaginary and the two rates are the same. 
In the CP violating case one can define a rate asymmetry as follows:
\begin{equation}
\sigma(\mu^+\mu^-\to H^+W^-)-\sigma(\mu^+\mu^-\to H^-W^+)\over
\sigma(\mu^+\mu^-\to H^+W^-)+\sigma(\mu^+\mu^-\to H^-W^+)
\label{eq:asym}
\end{equation}
Although this is a measure of the magnitude of the CP violation,
analogous to the direct CP asymmetry in the partial widths
of B hadron decays, the difference in the rates ($\sigma_{diff}$) 
is of more use experimentally: 
\begin{equation}
\sigma_{diff}=\sigma(\mu^+\mu^-\to H^+W^-)-\sigma(\mu^+\mu^-\to H^-W^+).
\end{equation}

\section{Numerical results}
We will present results for the CP violating 2HDM. For the CP conserving
2HDM, $\sigma_{tot}$ is usually very close in value to that of the CP
violating case (for the same choice of Higgs masses and $\tan\beta$),
and so we do not explicitly show results. The mass splittings 
of the Higgs bosons contribute to the $\rho$ parameter at the 1-loop level, 
and these extra contributions are constrained by 
$-0.0017\le \Delta\rho \le 0.0027$ \cite{Lang}. Therefore in our numerical
analysis we impose the formulae for $\Delta\rho$ in \cite{Rho}, 
which are valid for the CP violating 2HDM. 
We will assume integrated luminosities of the order 50 fb$^{-1}$ per year.

Measurements of $b\to s\gamma$ strongly restricts the allowed 
values of $M_{H^\pm}$ in the 2HDM with Model II type couplings.
Recent measurement suggest $M_{H^\pm}\ge 200$ GeV for $\tan\beta\ge 1$
\cite{Bor}.  

We show in Fig.~2a and 2b $\sigma_{tot}$ and $\sigma_{diff}$ 
as a function of $\sqrt s$, for 
$\tan\beta=4,20,50$. We have fixed the Euler angles 
such that the values $O_{21}=O_{22}=O_{23}=1/\sqrt 3$ are reproduced, and the
masses of $H_1,H_2,H_3$ are fixed at $100,400,700$ GeV respectively;
we also take $M_{H^\pm}=200$ GeV.
In Fig.~2a one can clearly see the large rises in $\sigma_{tot}$ 
when $\sqrt s\approx 
M_{H_i}$, which corresponds to the familiar resonance effect.
Such an enhancement is never possible in the MSSM case \cite{AAD} since
$M_A\approx M_H\approx M_{H^\pm}$, and so the conditions for on-shell
production ($\sqrt s\ge M_{H^\pm}+M_W$), and the resonance condition
($\sqrt s=M_{H_i}$)
can never simultaneously be satisfied. Fig.~2a shows 
that $\sigma_{tot}$ is maximized at the resonance
($\sqrt s\approx M_{H_i}$) and large $\tan\beta$. In such cases
$\sigma_{tot}\ge 1000$ fb is possible, and represents a copious
source of $H^{\pm}$.

In Fig.~1b we can see that $\sigma_{diff}$ is maximized with the same conditions
that maximized $\sigma_{tot}$,
and is always negative for the input parameters considered. 
Values of  $\sigma_{diff}$ up to 150 fb are possible for large $\tan\beta$.
With the expected luminosities of order 50 fb$^{-1}$, even 
$\sigma_{diff}\ge 2$ fb would lead to a mismatch of 
$\ge 100$ events in the rates for $H^+W^-$ and $H^-W^+$. 

In Fig.~3 we fix two Higgs masses almost equal ($M_{H_2}=400$ GeV,
$M_{H_3}=410$ GeV), and show the dependence 
of $\sigma_{tot}$ and  $\sigma_{diff}$ on $\sqrt s$.
We take $M_{H_1}=280$ GeV, $\tan\beta=50$ and the Euler angles
are the same as in Fig.2.   
In this case the cross-sections are strongly peaked at the
resonance, where $\sigma_{tot}\approx 4700$ fb and $\sigma_{diff}\approx 
1200$ fb, and this corresponds to an asymmetry (eq.(\ref{eq:asym})) of 
$\approx -26\%$. Away from resonance the cross-sections fall sharply with 
$\sqrt s$, in contrast to the case in Fig.~2a and 2b where sizeable
values for  $\sigma_{tot}$ and  $\sigma_{diff}$ were possible over a 
wide range of $\sqrt s$.

In Fig.~4a we plot $\sigma_{diff}$ as a function of $O_{23}$, for the same 
Higgs mass input parameters as used in Fig.~2. We fix $\tan\beta=50$ and
$\sqrt s=400$ GeV, and vary 2 Euler angles in order to explicitly
show the dependence of $\sigma_{diff}$ on $O_{ij}$.
One can see that the maximum value of $\sigma_{diff}$ arises
when $O_{23}=1/\sqrt 3$, which is the maximum CP violation 
condition applied in Fig.~2. Note that $\sigma_{diff}$
may be both positive and negative. The inner dots are eliminated 
by the $\rho$ parameter constraint, while the thicker dots
survive. We note that the latter points include the points that maximally
violate CP.  The $\rho$ parameter constraint has a strong effect on 
the magnitude of $\sigma_{tot}$, and rules out a sizeable parameter 
space where $\sigma_{tot}$ exceeds 3000 fb.
This is shown is Fig.~4b, where one can see that the points which
correspond to the largest values of $\sigma_{tot}$ are eliminated
by the $\rho$ parameter constraint.

\section{Conclusions}
We have considered the mechanism $\mu^+\mu^-\to H^\pm W^\mp$ in the
context of the CP violating 2HDM, which proceeds via 
Higgs mediated s-channel diagrams and $\nu_{\mu}$ exchange in
the t-channel. We showed that large values are
possible for both the total cross-section ($\sigma_{tot}$) and the 
difference in the cross-sections ($\sigma_{diff}$) for $H^+W^-$ and 
$H^-W^+$. The latter
represents a novel way of probing CP violating effects in the
Higgs sectors. The CP violation originates from the interference
between the weak phases in the vertices ($H_iH^{\pm}W^{\mp},
H_i\bar{\mu}\mu$) and the strong phases in the propagators.
We showed that both $\sigma_{tot}$ and $\sigma_{diff}$ are maximized
for large $\tan\beta$ and for $\sqrt s\approx M_{H_i}$, the latter
corresponding to the familiar resonance effect. 
Values of $\sigma_{tot}\ge 4000$ fb are possible at resonance 
for $\tan\beta=50$, and this provides a copious source of $H^{\pm}$.
Large values of $\sigma_{diff}$ provide a clear way of observing
CP violation, and we found that $\sigma_{diff}\ge 1000$ fb is possible.
Even $\sigma_{diff}\ge 2$ fb would correspond to a mismatch of $\ge 100$
in the number of $H^+W^-$ and $H^-W^+$ events, which should be readily 
observable.

\section*{Acknowledgements}  
A.G. Akeroyd was supported by the Japan Society for Promotion of Science (JSPS).
S. Baek was supported by Korea Science and Engineering Foundation (KOSEF).
We thank C. Dove for reading the manuscript.

\newpage
\renewcommand{\theequation}{C.\arabic{equation}}
\setcounter{equation}{0}

\subsection*{Figure Captions}

\vspace*{0.5cm}

\begin{itemize}
\item[Fig.2a]
$\sigma_{tot}$ as a function of $\sqrt s$ for various values of $\tan\beta$.
We take $M_{H_1}=100$ GeV, $M_{H_2}=400$ GeV, $M_{H_3}=700$ GeV.
\item[Fig.2b]
$\sigma_{diff}$ as a function of $\sqrt s$ for various values of $\tan\beta$.
For $M_{H_i}$ we use the values in Fig.2a.
\item[Fig.3\,\,\,]
$\sigma_{tot}$ and $\sigma_{diff}$ as a function of $\sqrt s$.
We take $\tan\beta=50$ and $M_{H_i}$ as displayed in the figure. 
\item[Fig.4a]
$\sigma_{diff}$ as a function of $O_{23}$. We fix $\tan\beta=50$
and $M_{H_i}$ are the same as in Fig.2a. The thin dots violate the
$\rho$ parameter constraint.
\item[Fig.4b]
Same as Fig.4a but for $\sigma_{tot}$.
\end{itemize}

\newpage
\renewcommand{\thepage}{}
\begin{minipage}[t]{19.cm}
\setlength{\unitlength}{1.in}
\begin{picture}(0.1,0.1)(0.8,8.2)
\centerline{\epsffile{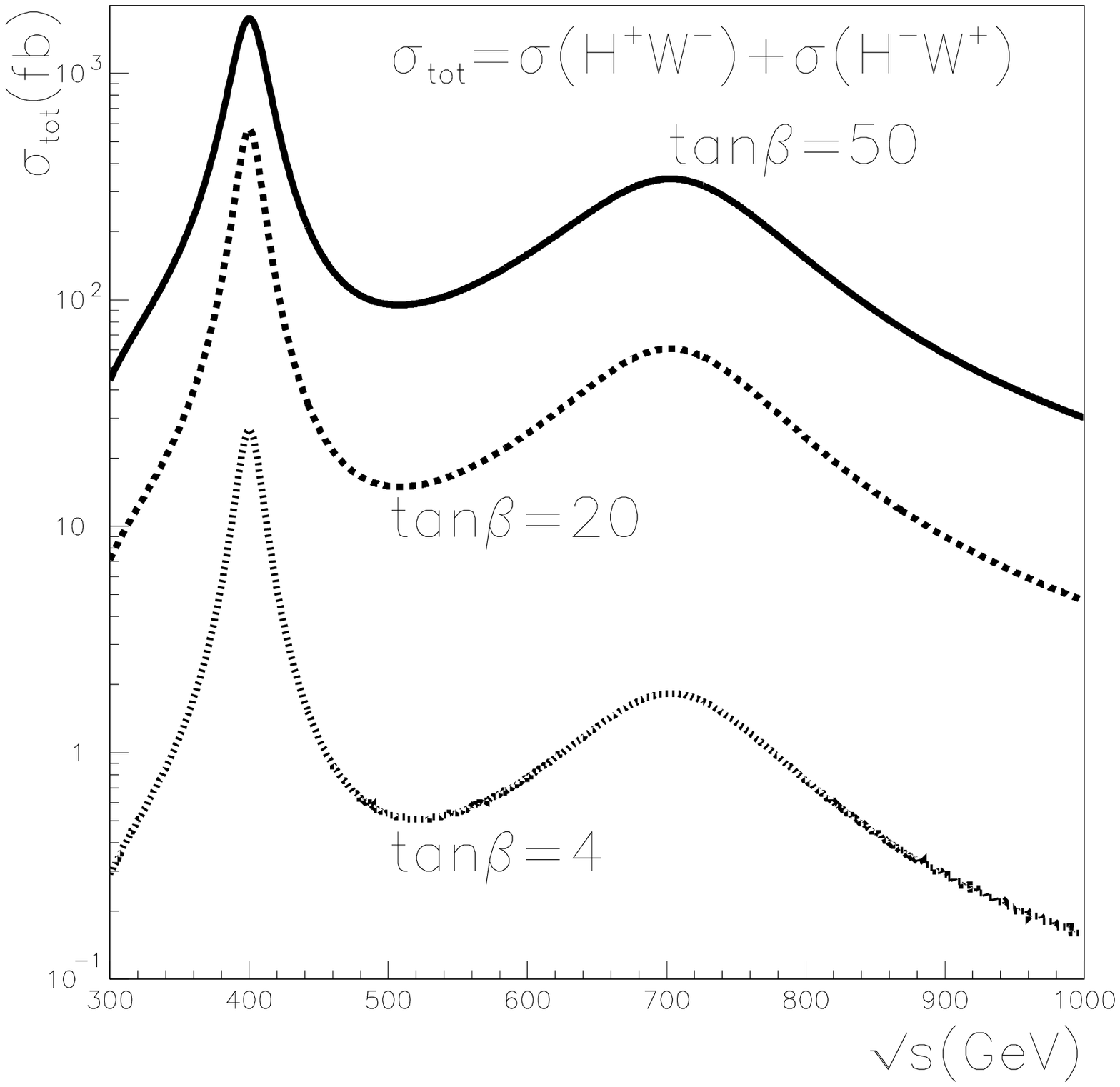}}
\end{picture}
\end{minipage}
\centerline{\bf{Figure. 2a}}

\newpage
\begin{minipage}[t]{19.cm}
\setlength{\unitlength}{1.in}
\begin{picture}(0.1,0.1)(0.8,8.2)
\centerline{\epsffile{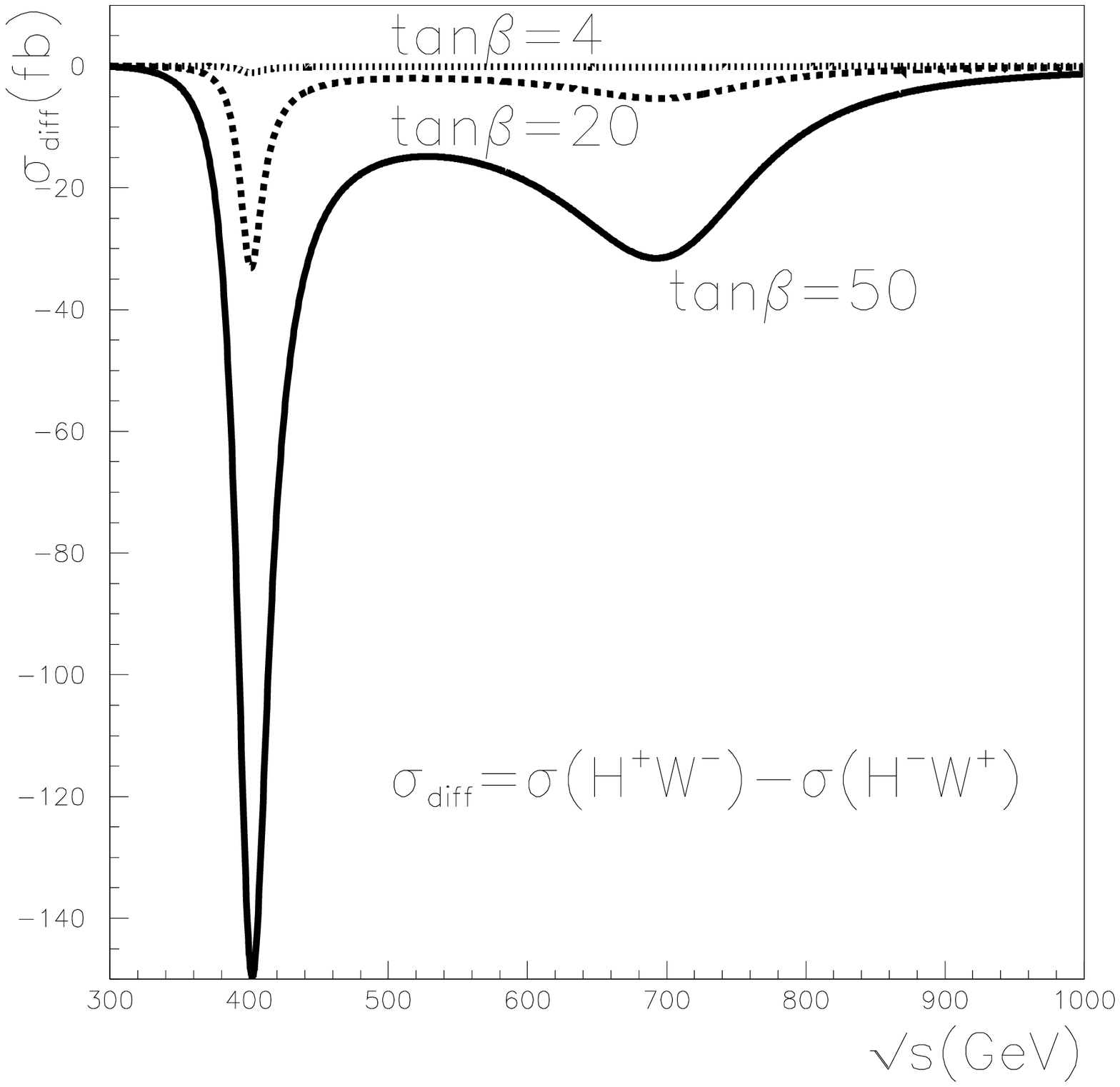}}
\end{picture}
\end{minipage}
$\ $
\vspace{5cm}
\centerline{\bf{Figure. 2b}}

\newpage
\begin{minipage}[t]{19.cm}
\setlength{\unitlength}{1.in}
\begin{picture}(0.1,0.1)(0.8,8.2)
\centerline{\epsffile{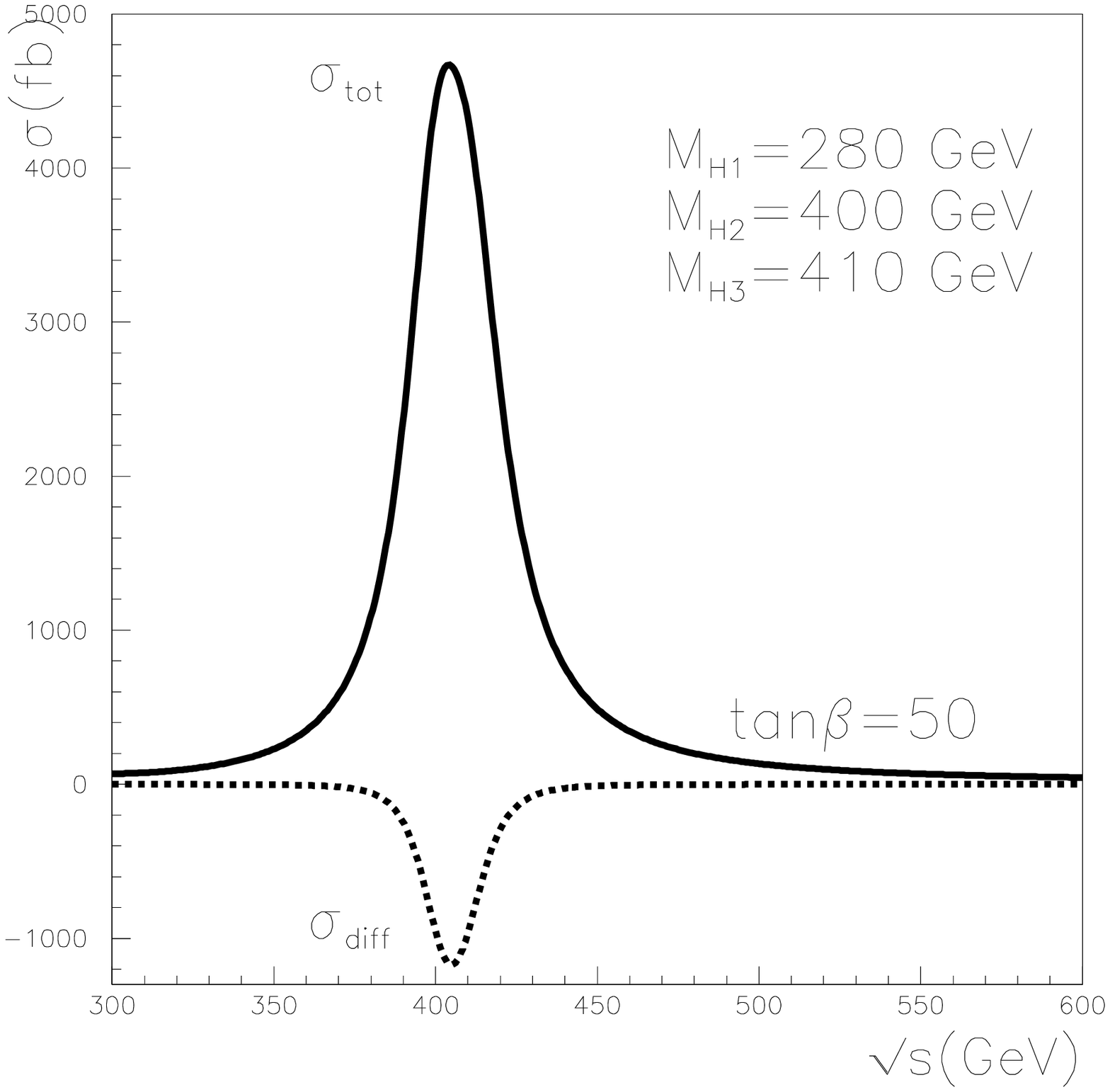}}
\end{picture}
\end{minipage}
\centerline{\bf{Figure. 3}}

\newpage
\begin{minipage}[t]{19.cm}
\setlength{\unitlength}{1.in}
\begin{picture}(0.1,0.1)(0.8,8.2)
\centerline{\epsffile{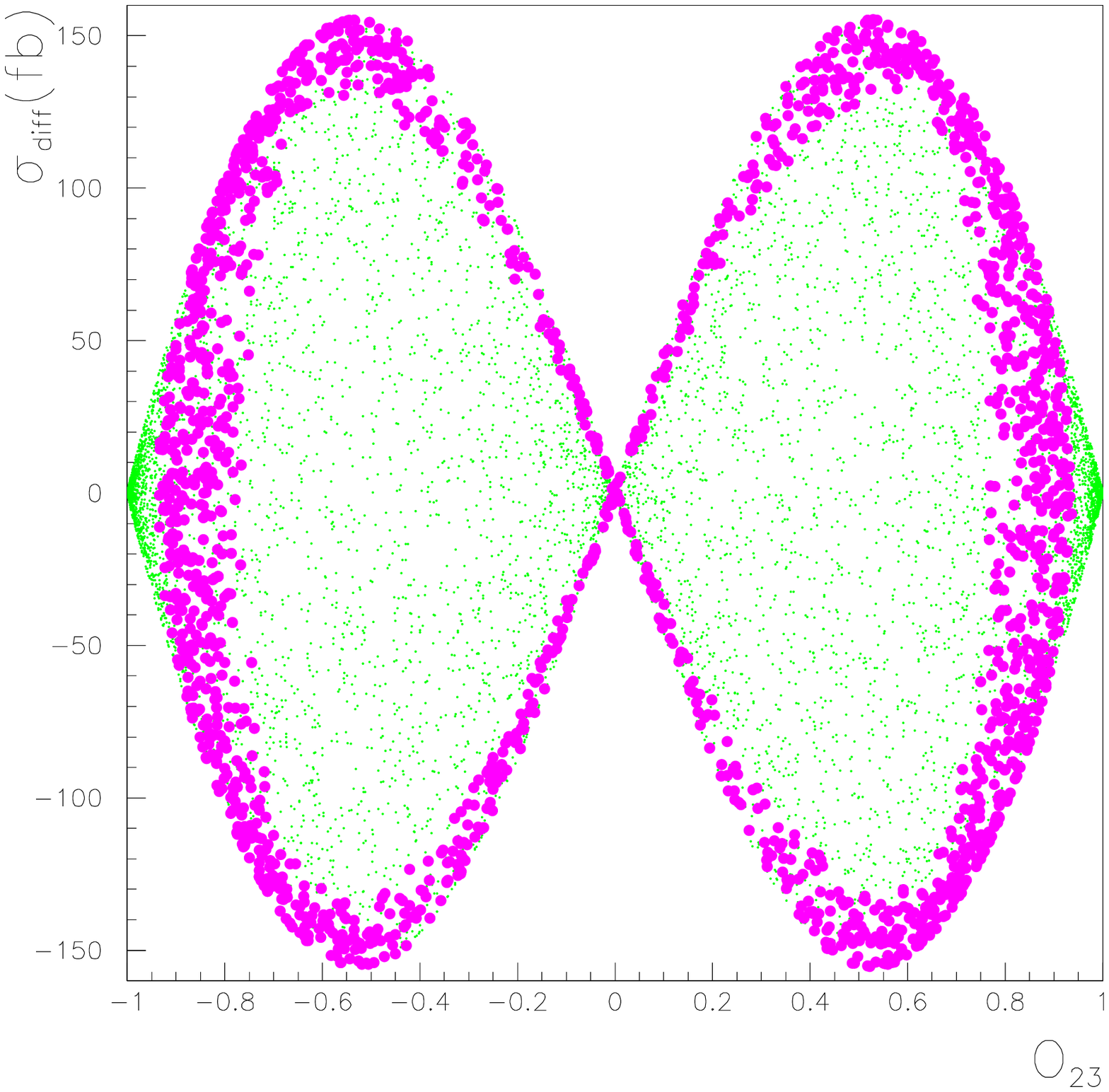}}
\end{picture}
\end{minipage}
\centerline{\bf{Figure. 4a}}

\newpage
\begin{minipage}[t]{19.cm}
\setlength{\unitlength}{1.in}
\begin{picture}(0.1,0.1)(0.8,8.2)
\centerline{\epsffile{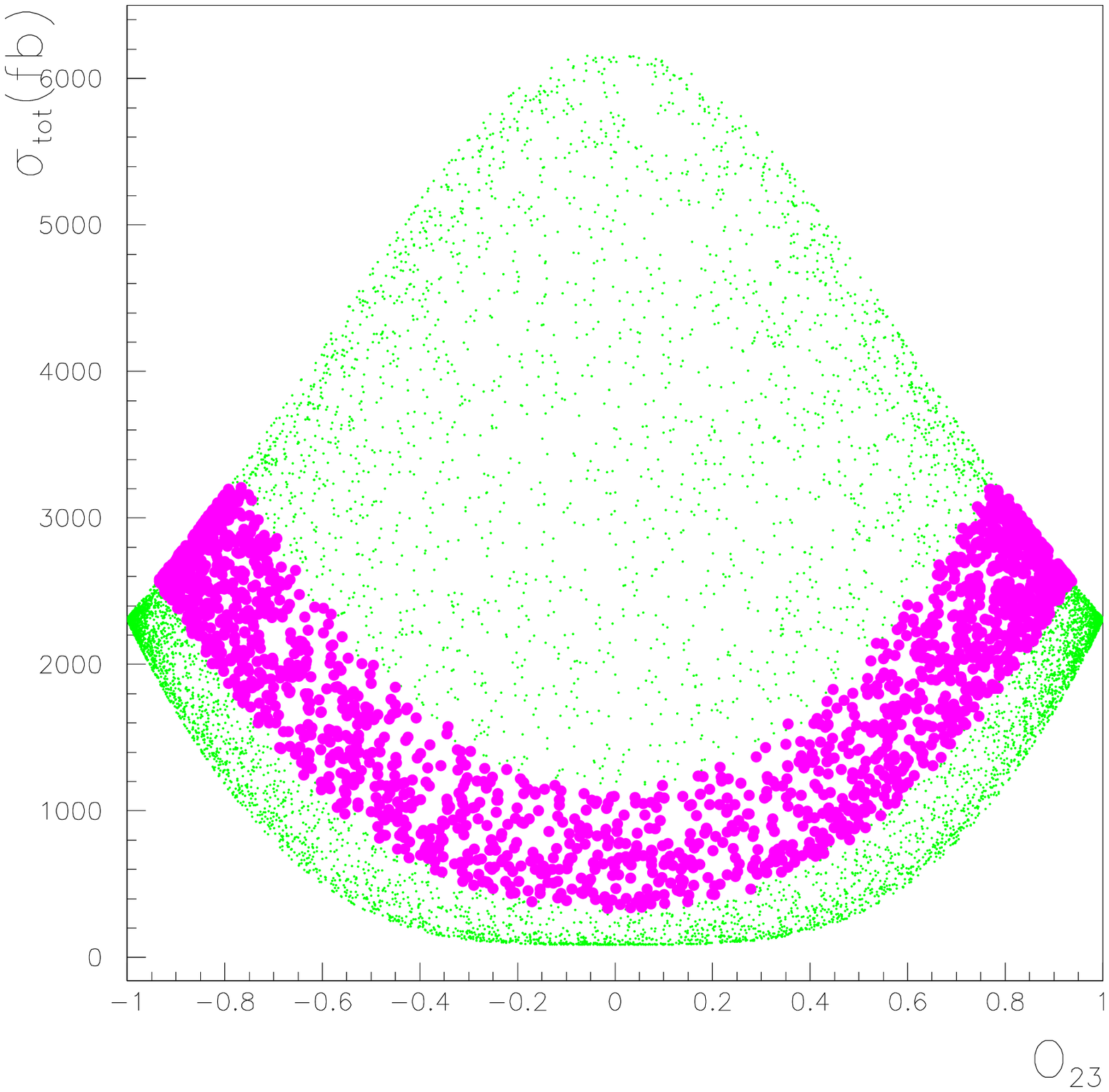}}
\end{picture}
\end{minipage}
\centerline{\bf{Figure. 4b}}

%
%
%
\end{document}